\documentclass[floatfix,showpacs,twocolumn,amssymb,amsfonts,amsmath,aps]{revtex4}
\usepackage{graphicx,epsfig}

\usepackage{latexsym}
\usepackage{graphicx}

\begin{document}
\title{Solidified Fillings of Nanopores}
\author{Patrick Huber}
\email[E-mail: ]{p.huber@physik.uni-saarland.de}
\affiliation{Technische Physik, Universit\"at des Saarlandes,
66041 Saarbr\"ucken, Germany}

\author{Klaus Knorr}
\affiliation{Technische Physik, Universit\"at des Saarlandes,
66041 Saarbr\"ucken, Germany}
\date{\today}

\begin{abstract}

We present a selection of x-ray and neutron diffraction patterns
of spherical (He, Ar), dumbbell- (N$_{2}$, CO), and chain-like
molecules (n-C$_{9}$H$_{20}$, n-C$_{19}$H$_{40})$ solidified in
nanopores of silica glass (mean pore diameter 7nm). These patterns
allow us to demonstrate how key principles governing
crystallization have to be adapted in order to accomplish
solidification in restricted geometries. $^{4}$He, Ar, and the
spherical close packed phases of CO and N$_{2}$ adjust to the pore
geometry by introducing a sizeable amount of stacking faults. For
the pore solidified, medium-length chain-like n-C$_{19}$H$_{40}$
we observe a close packed structure without lamellar ordering,
whereas for the short-chain C$_{9}$H$_{20}$ the layering principle
survives, albeit in a modified fashion compared to the bulk phase.
\end{abstract}
\maketitle

\section{Introduction}

Simple geometric considerations along with the goal of minimizing
the free energy of a system allow one to derive key principles of
the microscopic architecture of crystalline, condensed matter,
among them the close packing principle \cite{Buerger1963}. As the
structures of simple van-der-Waals molecular crystals testify, the
detailed manifestations of these building principles depend
sensitively on the symmetry and the interaction of the basic
building blocks (atoms, molecules or macromolecules)
\cite{Klein1977}. Here, we would like to give some flavor which
key crystallization principles survive or how they have to be
altered in order to allow a system to solidify in extreme spatial
confinement, that is in a geometry which is restricted, at least
in one direction, on the order of the size of its building blocks.
Our conclusions are drawn from the study of the structure of Ar,
N$_{2}$, CO, and two n-alkanes in the nanopores of silica glass.
We will present the structure of the confined molecular crystals
and discuss them with respect to the basic building principles
established in the corresponding bulk phases. Some attention will
also be paid on structural solid-solid phase transitions within
the confined crystalline phases.

By filling fraction dependent measurements both on the structure
 \cite{Huber1999} and on the dynamics \cite{Baumert2002} of
van-der-Waals systems embedded in porous glass one can show that
the pore condensates can usually be decomposed in two components:
The first two or three monolayers close to the pore walls form a
disordered, amorphous phase and at least for the silica matrices,
no melting or freezing transitions have been observed for this
component. Thus, one can term them as ``dead`` monolayers. By
contrast, the second part of the pore filling, located closer to
the pore center, leads a ``life of its own'' characterized by a
structure and thermodynamics which are reminiscent of the bulk
behavior, at least for pore diameters of the order of 10nm. In the
following, we shall focus on the structure of this second part and
will refer to it as \textit{pore solid}.

\section{Experimental}

As hosts we have chosen either Vycor glass or a controlled pore
glass (``Gelsil'' from Geltech, Orlando, FL). Both substrates are
practically pure fused silica. The structure of the nanopores of
both matrices can be described as a network of 3D randomly
oriented, connected pores with relatively uniform diameter d $\sim
$ 7nm \cite{Levitz1991}. The x-ray diffraction patterns have been
recorded using a Bragg-Brentano para-focussing geometry. For more
details on the experimental setup we would like to refer to ref.
\cite{Huber1999}. The elastic neutron diffraction experiment on
$^{4}$He has been carried out on the 2-axis diffractometer D20 of
the Insitut Laue Langevin, Grenoble (France). Both the neutron as
well as the x-ray diffraction patterns will be presented as plots
of the scattered intensity versus the modulus of the scattering
vector $q$, $q=4\pi $/$\lambda $ sin($\Theta )$, where $\lambda $
corresponds to the wavelength of the x-rays and neutrons, resp.
Additionally, the scattering angle 2$\Theta $ is plotted on the
top axis for the x-ray diffraction patterns ($\lambda $=1.542
{\AA}$^{-1})$.

\section{Results}

\begin{figure*}[!]
\epsfig{file=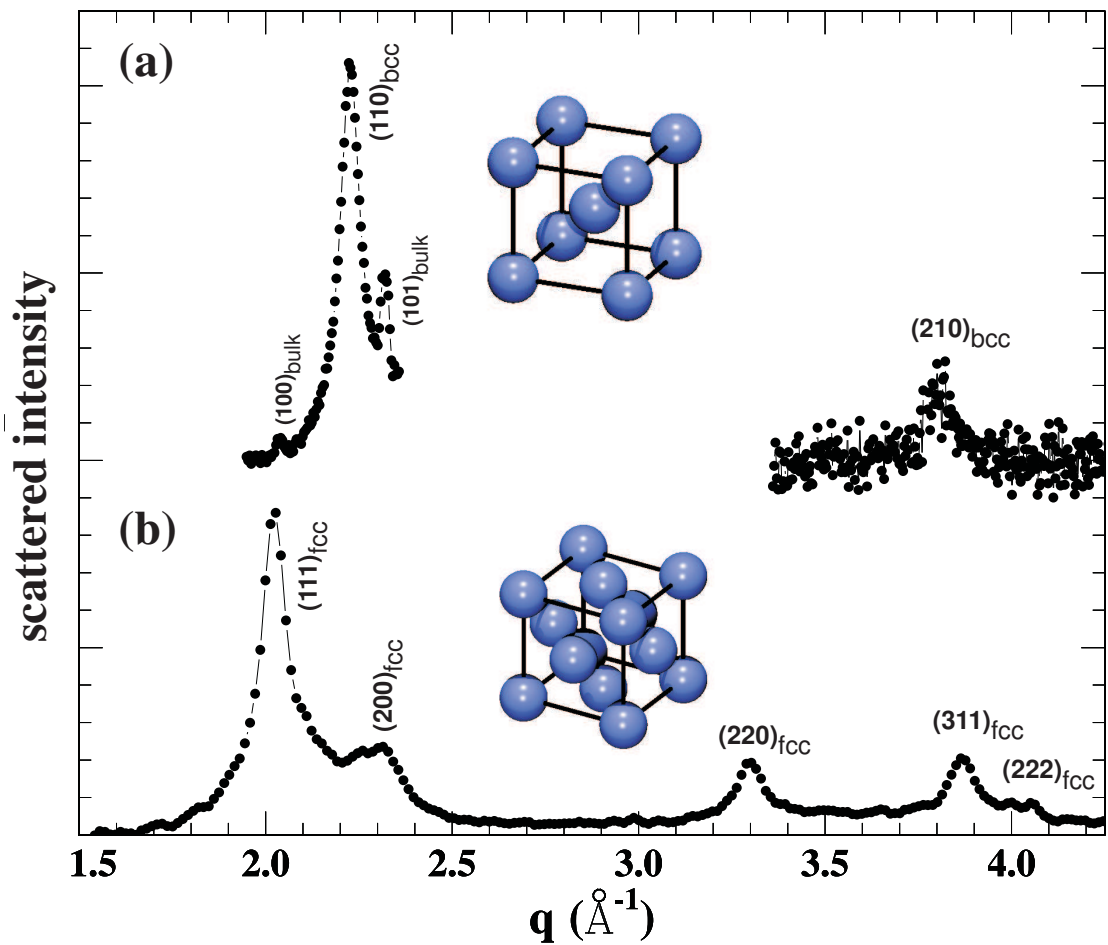, angle=0, width=1.7\columnwidth}
\caption{\label{fig1} (a) Neutron diffraction pattern of pore
condensed $^{4}$He at p=70bar, T=1K \cite{Wallacher2005}. The
reflections (hkl) are indexed on the basis of a bcc cell. The
cartoon depicts the bcc structure. Note, the intensity of the
(210) bcc reflection has been multiplied by a factor of 50. (b)
X-ray diffraction pattern of pore condensed Ar at T=50K
\cite{Huber1999}; indicated are the fcc (hkl) reflections. As
inset one can find an illustration of the fcc structure.}
\end{figure*}

\begin{figure*}[!]
\epsfig{file=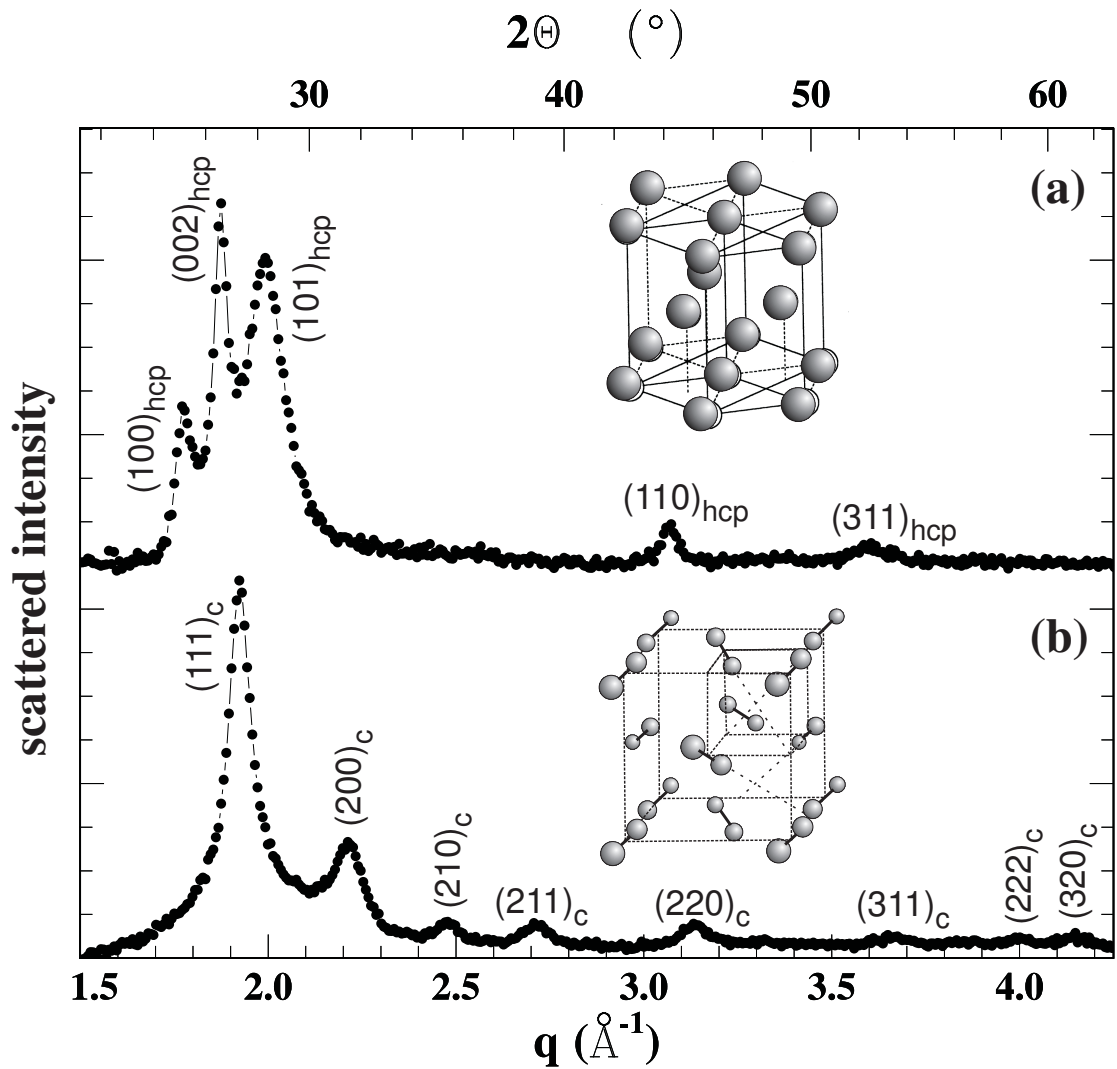, angle=0, width=1.7\columnwidth}
\caption{\label{fig2}(a) X-ray diffraction pattern of N$_{2}$
confined in GelSil at T=50K. The reflections are indexed on the
basis of a hcp lattice \cite{Huber1999b} (b) X-ray diffraction
pattern of pore condensed CO at T=50K; indicated are the Pa3 (hkl)
reflections \cite{Huber1999b}. The insets in panel (a) and (b)
illustrate the hcp and Pa3 structures, respectively.}
\end{figure*}

\begin{figure*}[!]
\epsfig{file=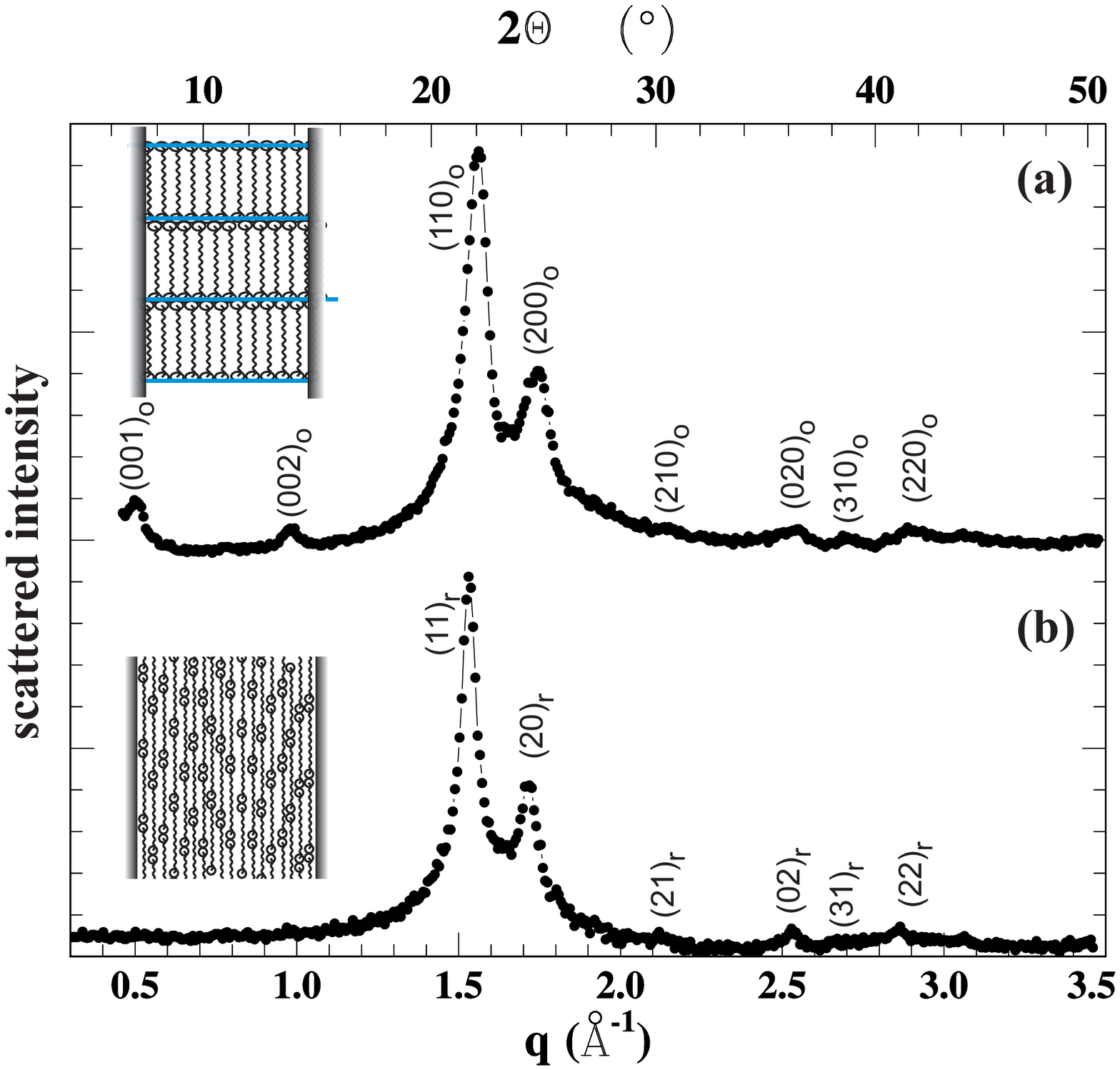, angle=0, width=1.7\columnwidth}
\caption{\label{fig3}(a) X-ray diffraction patterns of
n-C$_{9}$H$_{20}$ in GelSil70 at T=180K. The reflections are
indexed assuming an orthorhombic unit cell. The inset depicts an
illustration of the structure. The molecules' long axes are
aligned parallel to the pores and obey a lamellar ordering. (b)
X-ray diffraction pattern of pore condensed n-C$_{19}$H$_{40}$ at
T=200K. The reflections are indexed on the basis of a
two-dimensional rectangular mesh. The cartoon depicts the
``nematocrystalline'' state of the pore solid \cite{Huber2004}.}
\end{figure*}

\subsection{Helium and Argon}

For $^{4}$He the quantum mechanical zero point energy is so large
compared to the normal attractive interactions that it exists as
solid only under pressure $p$ ($p>$25bar), and then with very
small cohesive energies, large molar volumes and large
compressibilities \cite{Klein1977}. Bulk $^{4}$He forms either a
hexagonal close packed (hcp) or a body centered cubic (bcc) phase.
The more loosely packed bcc-structure occurs at higher
temperatures and is believed to be stabilized relative to the hcp
structure for entropic reasons.

A diffraction pattern of confined $^{4}$He (T=1K, p=70bar) is
depicted in Fig 1. (a) \cite{Wallacher2005}. The part of the
pattern containing reflections from the sample cell walls and the
various thermal shields ($q$-range from 2.4-3.5{\AA}$^{-1})$ has
been removed. The scattering intensity is dominated by one peak at
$q=2.25 {\AA}^{-1}$ that can be attributed to confined He, whereas
the sharp peaks on the left and on the right wing of this peak are
due to hcp bulk crystallites, sitting outside the pore space.
Additionally a broad, weak peak at $q=3.75{\AA}^{-1}$ can be
identified as belonging to confined $^{4}$He. A detailed analysis
of the temperature dependence of the diffraction pattern that
considers different packing sequences of triangular net planes
(ABC stacking $\Leftrightarrow$ fcc, AB stacking $\Leftrightarrow$
hcp, random stacking) and the bcc structure, identifies the two
observed peaks as the (110) and (210) bcc reflections. The
intrinsic width $\Delta q$ of the two peaks indicate a coherence
length L of the $^{4}$He crystallites of about 10nm. L is slightly
larger than the pore diameter. The two Bragg peaks represent only
about 30{\%} of the pore filling, whereas the non-crystalline
fraction amounts to about 70{\%}. Additionally to two ``dead''
monolayers of immobile molecules next to the pore walls the
thermodynamic path (low T, high p), which one has to take in order
to solidify $^{4}$He in the pores leads to a liquid-like shell
between the pore solid in the center of the pores and the
amorphous wall coating. Further cooling below 1K leads to no
qualitative change of the diffraction pattern. Quite different to
the situation for bulk He, the loosely packed bcc structure in the
pore center is stable down to lowest temperatures. Thus, the
bcc-hcp lattice reconstruction, known from bulk $^{4}$He, is
suppressed in the pores.

In contrast to the quantum crystal $^{4}$He the zero point motion
of the atoms of solid Ar are negligible rendering it to one of the
most simple, classical crystalline solids \cite{Klein1977}. Kept
under its own vapor pressure, Ar solidifies into a fcc crystalline
structure upon cooling below its bulk triple point of
$T_{3}^{bulk}$=83.8K. Temperature dependent diffraction patterns
of Ar confined in Vycor start to exhibit Bragg peaks upon cooling
below 76K, only. One observes a reduced freezing temperature in
the pores. The diffraction pattern of the pore solid, as it is
shown in Fig. 1(b) for T=60K, can be indexed by referring to the
bulk fcc Ar structure. In the accessible q-range one can identify
the (111), (200), (220), (311) and (222) reflections. Apart from
the finite size broadening, the width of the fcc reflections show
a variation characteristic of crystals suffering from a
substantial amount of stacking faults. If $\alpha $ is the
probability that an AB stack of triangular (111) planes is
followed by a plane in the A position (hcp-like) rather than in
the C position (fcc-like) and L is a characteristic crystallite
size, one arrives at a satisfactory description of the diffraction
pattern of the pore solid with L=100nm and $\alpha $=0.055. The
value of L exceeds the pore diameter by far, which points again to
a rather large coherence length of the crystallites along the
pores.

\subsection{Nitrogen and Carbonmonoxide}

N$_{2}$ and CO are molecules of similar size and shape. As bulk
systems, they have closely related structural and thermodynamical
properties. The melting temperatures are almost equal:
T$_{3}^{bulk}$ (N$_{2})$=63.15K, T$_{3}^{bulk}$ (CO)=68.9K. Bulk
N$_{2}$ and CO are isomorphous in both solid phases. The high-T
$\beta {\rm g}$hase is hcp with a practically ideal c/a ratio --
see illustration in Fig. 2(a). Here the molecules are
orientationally disordered due to rapid reorientations. At lower
temperatures the bulk solids undergo a first-order phase
transition into the $\alpha $ phase (cubic, Pa3), which involves a
change of the center-of-mass lattice from the hcp-stacking
sequence AB{\ldots} to the fcc-stacking sequence ABC{\ldots} and
an orientational ordering of the molecular axes, i.e. of the
quadrupole moments, such that each of the four molecules of the
cell points along one of the four $<$111$>$ directions -- compare
illustration in Fig. 2(b). This orientation of the molecules
appears as a compromise of the cubic symmetry of the center of
mass lattice and the quadrupolar part of the interaction of the
molecules: The resulting mutual orientation of the long axes of
the molecules is given by the tetrahedral angle of 109 degrees,
which is reasonably close to an angle of 90 degrees minimizing
their quadrupolar interaction. Solidification of N$_{2}$ in
GelSil, indicated by the appearance of the hcp Bragg peaks, occurs
at T=54K. The diffraction pattern of confined N$_{2}$ shows five
Bragg peaks, which all coincide with reflections of bulk hcp
$\beta $-N$_{2}$, although with altered intensity ratios and
enhanced linewidths compared to the pattern of the bulk phase. The
(102) reflection is missing. The (103) reflection is extremely
broadened, but can be still identified. By contrast, the (002)
reflection dominates the diffraction pattern and is remarkably
sharp. The variation of the intrinsic peak width $\Delta $q as a
function of the crystallographic direction can be described by a
preferred growth of the crystallites with their c-axes along the
pores and a sizeable amount of stacking faults in the pore solid.
This conclusion is corroborated by diffraction studies varying the
pore diameter as well as studies of N$_{2}$ in well-aligned pores
of porous silicon \cite{Hofmann2005}. Thus, pore confined nitrogen
establishes a structure very related to the bulk $\beta $ phase.
Upon cooling, however, no transition into the orientational
ordered $\alpha {\rm g}$hase has been observed. This solid-solid
transition could be investigated for pore confined CO, only. It
forms the same hcp phase as just presented for confined N$_{2}$,
however, at T=54K the center of mass lattice rearranges according
to the $\beta {\rm t}\alpha $ phase transition scenario and for
T$<$54K one observes a diffraction pattern typical of the cubic
Pa3 structure -- compare Fig. 2(b). As a peculiarity of confined
CO the occurrence of a fcc phase upon heating has been found. The
orientational order of the molecules is lost while maintaining the
cubic lattice.

\subsection{Nonane and Nonadecane}

In the following we focus on the structure of more complex molecular
crystals, built out of chain-like molecules, where the length l of the
rectified molecules is of the order of the pore diameter. We have chosen the
linear hydrocarbons n-C$_{9}$H$_{20}$ and n-C$_{19}$H$_{40 }^{ }$with
l=1.3nm and l=2.6nm, resp. The C atoms of the zigzag backbone of these
linear hydrocarbons are all in the trans-configuration, so all of them are
located in a plane. The molecular crystals form layered structures. For
n-C$_{19}$H$_{40}$ the molecules are aligned perpendicular to the layers.
Within the layers the molecules are close packed, side by side, in a
quasi-two-dimensional array. For low temperatures the azimuth of the
rotation of the --C-C- plane around the long z-axis of the molecules
alternates in a herringbone fashion. In the short-length n-C$_{9}$H$_{20}$
the molecules are tilted by about 15deg with respect to the layer normals.

The diffraction pattern of confined n-C$_{9}$H$_{20}$ -- see Fig.
3 -- indicates a layering of the molecules (two peaks at low q)
and additionally a herringbone type ordering of the molecules
within the layers (6 peaks at q$>$1.4 {\AA}$^{-1})$. The
reflections at higher q can be indexed as (hk0) in terms of an
orthorhombic unit cell. This means that any correlations of the
lateral positions of the layers are lost in the pores. If such
correlations existed, peaks of the type (hkl), l$<>$0 should show
up. Moreover, the layering distance extracted from the (001)$_{o}$
and (002)$_{o}$ peaks at low q agree with the length of the
rectified molecules, thus the confinement imposes an alignment of
n-C$_{9}$H$_{20}$ parallel to the layer normal. Presumably, the
layer normals are oriented parallel to the pore axis, which
results in a structure reminiscent of the crystal-E phase of
rod-like liquid crystals \cite{Kumar2001}.

As can be seen by a comparison of Fig. 3(a) and (b) the
diffraction pattern of n-C$_{19}$H$_{40}$ is identical to the one
of n-C$_{9}$H$_{20}$ except for the missing (00l) layering peaks
at low q (q$<$1.4 {\AA}$^{-1})$. The resulting structure of
confined n-C$_{19}$H$_{40}$ is depicted in Fig. 3(b): The
molecules' position along the pore axes have no correlations. One
basic ordering principle of the bulk crystalline alkane, i.e. the
layering, is suppressed, rendering the system effectively
two-dimensional (2D) \cite{Huber2004}. Moreover, an analysis of
the coherence length of the 2D crystals based on the width of the
(hk)$_{r}$ reflections yields 7.5nm. This value is close to the
mean pore diameter, which suggests an alignment of the molecules
along the pore axis. Isomorphous structures, termed
``nematocrystalline'', have been reported for the bulk phases of
natural waxes, e.g. bee wax, consisting of complicated mixtures of
chain-like molecules, among them alkanes \cite{Dorset1999}.

\section{Summary}

Spherical close-packed structures can accommodate to the pore geometry by
introducing of a sizeable amount of stacking faults. The structures of the
confined chain-like molecules is affected in a more drastic way, in
particular the anisotropic pore structure calls for an alignment of the
molecules parallel to the pore axes. Nevertheless, the structures follow the
close packing principles; even the medium-length n-C$_{19}$H$_{40}$ pursues
this concept by establishing a ``nematocrystalline'' state.

Extreme cases of structural disorder such as random-stacking or
random close packing have not been found for the pore solids
investigated. The preference to form structures with higher
symmetry obeying subtleties of the symmetry of the molecules'
interaction potentials, which lead to the fcc-Ar, hcp- or bcc-
N$_{2}$, CO, and $^{4}$He, resp. and finally to the
herringbone-type ordering of the linear hydrocarbon chains,
prevails even for the crystallization in nano-confinement. If one
considers the distribution of the pore diameters and the
tortuosity of the pores, in general all deviations of the employed
porous hosts from an idealized pore space which presumably
additionally favor disordered structures, our findings testify
even more to the robustness of the basic crystallization
principles. As witnessed by the suppression of the bcc-hcp
transition in confined He, the absence of the $\alpha {\rm t}\beta
$ transition in confined N$_{2}$, and the occurrence of a fcc-CO
phase in the pores, the phase sequence in the pores can
nevertheless be significantly affected by the confinement. For
pore condensates embedded in well-defined nanopores of template
grown MCM-41 or SBA-15 silica matrices \cite{temppores} quite
similar structures have been observed \cite{Morishige2000}. Thus,
the detailed geometry and topology of the matrix is of minor
importance for the crystallization; they are, however, crucial for
an understanding of the thermodynamical aspects of pore
condensates, e.g. the freezing/melting process or
condensation/sublimation phenomena in the pores
\cite{Christenson2001}.

Another interesting aspect concerns the thermodynamic stability of
pore condensates. There is agreement that capillary condensed
cryogenic liquids such as Ar, He, CO and N$_{2}$ are stable in
mesoporous glasses such as Vycor. In the solid regime, however,
several authors report changes that have been interpreted by a
migration of the filling out of the pores and have been referred
to as ``mobility transition'', a ``new type of structural phase
transition'', and ``dewetting transition'' \cite{Silva2002}. We
could not find any hints for such transitions of the cryosolids
investigated. Nevertheless, we would like to mention that we
gather increasing evidence for a thermodynamic metastability of
the solid phases of medium- and long-chain n-alkanes in an ongoing
study aimed at an understanding of the structure and
thermodynamics of hydrocarbons in nanopores.

Finally we would like to mention that studies on crystallization
or, more generally spoken, structural transitions in
nanoconfinement are not only interesting by their own virtue, but
also of relevance for such different phenomena as friction, frost
heave, transport through biomembranes, and crystal growth \cite{
Urback2004}. For instance, recently the immobilization and
crystallization of proteins in nanopores of Vycor and porous
silicon attracted attention in the flourishing field of structural
genomics \cite{Chayen2001}. It could be demonstrated that these
pore confined protein structures promote the formation of protein
bulk crystallites, one of the pivotal goals in this area of
science.

\acknowledgments{This work has been supported by the Deutsche
Forschungsgemeinschaft (SFB 277).}

\end{document}